\documentclass[ALICE,manyauthors]{cernphprep}
\usepackage{cite}
\newcommand{\jpsi}{\rm J/$\psi$}
\newcommand{\psip}{$\psi(\rm 2S)$}

\usepackage{rotating}
\usepackage{hyperref}
\usepackage[usenames,dvipsnames,svgnames,table]{xcolor}

\begin{document}%
%
%
\begin{titlepage}
\PHyear{2014}
\PHnumber{092}      
\PHdate{14 May}            
%
%
\title{Suppression of $\mathbf{\psi}$(2S) production \\ in \mbox{p-Pb} collisions at $\mathbf{\sqrt{{\textit s}_{\rm NN}}}$~=~5.02 TeV}

\ShortTitle{Suppression of $\mathbf{\psi}$(2S) production in \mbox{p-Pb} collisions at $\mathbf{\sqrt{{\textit s}_{\rm NN}}}$~=~5.02 TeV}
%
\Collaboration{ALICE Collaboration}%
\ShortAuthor{ALICE Collaboration}      
\begin{abstract}
The ALICE Collaboration has studied the inclusive production of the charmonium state $\psi(\rm 2S)$ in proton-lead (\mbox{p-Pb}) collisions
at the nucleon-nucleon centre of mass energy $\sqrt{s_{\rm NN}}$ = 5.02 TeV at the CERN LHC. The measurement was performed at forward ($2.03<y_{\rm cms}<3.53$) and backward ($-4.46<y_{\rm cms}<-2.96$) centre of
mass rapidities, studying the decays into muon pairs. In this paper, we present the inclusive production cross sections
$\sigma_{\psi(\rm 2S)}$, both integrated and as a function of the transverse momentum $p_{\rm T}$, for the two $y_{\rm cms}$
domains. The results are compared to those obtained for the 1S
vector state (J/$\psi$), by showing the ratios between
the production cross sections,
as well as the double ratios $[\sigma_{\psi(\rm 2S)}/\sigma_{\rm J/\psi}]_{\rm pPb}/[\sigma_{\psi(\rm 2S)}/\sigma_{\rm
J/\psi}]_{\rm pp}$ between \mbox{p-Pb} and proton-proton collisions. Finally, the nuclear modification factor for inclusive
$\psi(\rm 2S)$  is evaluated and compared to the measurement of the same quantity for J/$\psi$ and to theoretical models including
parton shadowing and coherent energy loss mechanisms. The results show a significantly larger suppression of the $\psi(\rm 2S)$ 
compared to that measured for J/$\psi$ and to models. These observations represent a clear indication for sizeable final state
effects on $\psi(\rm 2S)$  production.
\end{abstract}
\end{titlepage}
%
The physics of charmonia, bound states of the charm ($c$) and anti-charm ($\overline c$) quarks, is an
extremely broad and interesting field of investigation~\cite{Bra11}. The description of the various states
and the calculation of their production cross sections in hadronic collisions involve an interplay of
perturbative and non-perturbative aspects of Quantum ChromoDynamics (QCD)~\cite{Bod95}, which 
still today represent a significant challenge for theory~\cite{But11}. Charmonium
states can have smaller sizes than light hadrons
(down to a few tenths of a fm) and large binding
energies ($> 500$ MeV)~\cite{Eic80}. These properties make charmonia a useful probe of the hot nuclear matter
created in ultrarelativistic heavy-ion collisions, which can be seen as a plasma of deconfined quarks and
gluons (QGP) (see~\cite{QM12} for a recent overview of QGP studies). 
In particular, the $c\overline c$ binding can be screened by the high density of colour charges present in
the QGP, leading to a suppression of the yields of charmonia in high-energy nuclear collisions compared to
the corresponding production rates in elementary pp collisions at the same
energy~\cite{Mat86}. In the so-called ``sequential suppression'' scenario, the melting of a bound $c\overline
c$ state occurs when the temperature of the hot medium exceeds a threshold dissociation
temperature~\cite{Kar91,Dig01}, which depends on the binding energy of the state and can be calculated in lattice
QCD~\cite{Din12}. At LHC energies, where the number of produced $c\overline c$ pairs is large, this 
suppression effect can be partly counterbalanced by charmonium ``regeneration'' processes due to the recombination of charm quarks 
that occurs as the system cools and hadrons form~\cite{PBM00,The01,And07}.

Among the charmonium states, the strongly bound S-wave J/$\psi$ and the weakly bound radially excited \psip\
have received most attention in the context of QGP studies. Both decay to lepton pairs with a non-negligible
branching ratio (5.93\% and 0.77\%, respectively, for the $\mu^+\mu^-$ channel~\cite{Ber12}). The results
obtained by the NA50 collaboration at the CERN SPS showed a significant suppression of the J/$\psi$
 production in Pb-Pb collisions at $\sqrt{s_{\rm NN}}=17$ GeV~\cite{Ram05} and a comparatively larger suppression of the \psip~\cite{Ale07}, in qualitative agreement with sequential suppression models.
However, the same experiment also detected a significant suppression of both states (although not as strong as in \mbox{Pb-Pb}) in
proton-nucleus (\mbox{p-A}) collisions~\cite{Ale06}, where no QGP formation was expected. The same observation was made by other
fixed-target experiments studying \mbox{p-A} collisions at Fermilab 
(E866~\cite{Lei00}) and HERA (HERA-B~\cite{Abt07}). It was indeed realized that the charmonium yields are also sensitive 
to the presence of cold nuclear matter (CNM) in the target nucleus, and various mechanisms (nuclear parton 
shadowing~\cite{Esk09}, $c\overline c$ break-up via interaction with nucleons~\cite{Vog02,Kop01,Vog12}, 
initial/final state energy loss~\cite{Arl12}) were taken into account in order to describe experimental observations. In particular, these experiments observed a stronger suppression for \psip\ relative to J/$\psi$ at central rapidity, 
while at forward rapidity no difference was found within uncertainties. This feature of the results was interpreted in terms of pair 
break-up: at central rapidity the time spent by the $c\overline c$ state in the nuclear medium (crossing time) is typically larger
than the formation time of the resonances ($\sim 0.1$ fm/$c$~\cite{Arl00,McG13}), so that the  loosely bound \psip\ can be more
easily dissociated than the J/$\psi$. Conversely, in forward production the crossing time is smaller than the formation time and the influence of the nucleus on the pre-hadronic state is the same, independent of the particular resonance being produced~\cite{Vog00}.

More generally, the study of charmonia in \mbox{p-A} collisions can be used as a tool for a quantitative
investigation of the aforementioned processes, relevant in the context of studies of the strong interaction.
Therefore, measurements at high energies are important to test our understanding of the various mechanisms.
In particular, the pair break-up cross sections discussed above are expected to be strongly reduced due to
the increasingly shorter time spent by the $c\overline c$ pair in CNM. On the other hand, the other
effects listed above (shadowing, energy loss) are not expected to depend on the final quantum numbers of the
charmonium states. In such a situation, a similar suppression for the two charmonium states should be observed in high-energy 
\mbox{p-A} collisions.

In the context of comparative studies between the resonances, the PHENIX experiment at RHIC has recently published results
on the \psip\ suppression at central rapidity for \mbox{d-Au} collisions at $\sqrt{s_{\rm NN}} = 200$ GeV~\cite{Ada13}, by studying the nuclear modification factor $R^{\psi(\rm 2S)}_{\rm dAu}= 
{\rm d}N^{\psi(\rm 2S)}_{\rm dAu}/{\rm d}y/(N_{\rm coll}\times{\rm d}N^{\psi(\rm 2S)}_{\rm pp}/{\rm d}y)$,
which corresponds to the ratio of the production yields in \mbox{d-Au} and pp at the same energy, normalized
by the number of nucleon-nucleon collisions in \mbox{d-Au}. The ratio of the nuclear modification factors 
$R^{\psi(\rm 2S)}_{\rm dAu}/R^{\rm J/\psi}_{\rm dAu}$ is found to be smaller than 1, and
 strongly decreasing from peripheral to central \mbox{d-Au} events. The observation of a \psip\ suppression stronger than that of the \jpsi\ is in contrast to the expectation of a similar suppression as described above. 
Data from the LHC can be useful to shed further light on this observation, as nuclear crossing
times~\cite{McG13} may be as low as 10$^{-4}$ fm/$c$ for charmonium production at forward rapidity, implying a negligible influence of pair break-up processes and, in more general terms, to test our understanding of charmonium propagation in CNM. 

In this Letter, we present the first measurement of inclusive \psip\ production in $\sqrt{s_{\rm NN}}=5.02$ TeV 
\mbox{p-Pb} collisions at the LHC, carried out by the ALICE Collaboration, and we compare the results with those for 
J/$\psi$. The resonances were measured in the dimuon decay channel using the Muon Spectrometer (MS)~\cite{Aam11}, 
which covers the pseudorapidity range $-4<\eta_{\rm lab} <-2.5$. The other detectors involved in this analysis are: 
(i) the two innermost layers of the Inner Tracking System (Silicon Pixel Detectors, SPD), 
used for the determination of the primary vertex of the interaction and covering $|\eta_{\rm lab}|<2.0$ (first layer) 
and $|\eta_{\rm lab}|<1.4$ (second layer)~\cite{Aam10}; (ii) the two VZERO scintillator hodoscopes, used mainly for triggering purposes and covering  $-3.7<\eta_{\rm lab} <-1.7$ and 
$2.8<\eta_{\rm lab} <5.1$~\cite{Abb13}; (iii) the Zero Degree Calorimeters (ZDC), at 112.5 m from the interaction point~\cite{Abe12b}, used to remove collisions outside the nominal timing of the LHC bunches. Details 
of the ALICE experimental setup are provided elsewhere~\cite{Aam08}.

Due to the LHC design, the colliding beams have different energies per nucleon ($E_{\rm p} =
4$ TeV, $E_{\rm Pb} = 1.58\cdot A_{\rm Pb}$~TeV, where $A_{\rm Pb}=208$ is the mass number of the Pb nucleus). As a
consequence, the centre of mass of the nucleon-nucleon collision is shifted by $\Delta y = 0.465$ with
respect to the laboratory frame in the direction of the proton beam. Data were taken in two configurations, by inverting the sense
of the orbits of the two beams. In this way, both forward ($2.03<y_{\rm cms}<3.53$) and
backward ($-4.46<y_{\rm cms}<-2.96$) centre of mass rapidities were covered, with the
positive rapidity defined by the direction of the proton beam. We refer to the two data
samples as \mbox{p-Pb} and \mbox{Pb-p} respectively. The integrated luminosities for the two data samples are $L^{\rm pPb}_{\rm int}=5.01 \pm 0.19$ nb$^{-1}$ and $L^{\rm Pbp}_{\rm int}=5.81 \pm 0.20$ nb$^{-1}$~\cite{Gag13}.

Data were collected with a dimuon trigger, defined as the coincidence of the minimum-bias (MB) condition with the detection of two opposite-sign muon candidates in the trigger system of the MS. The MB condition is a coincidence between signals in the two VZERO hodoscopes and has $>99$\% efficiency for non-single diffractive events~\cite{Abe13b}. 
For the muon candidates, a transverse momentum $p_{\rm T,\mu}=0.5$ GeV/$c$ trigger threshold is applied. The effect of 
this threshold is not sharp, and the single muon trigger efficiency reaches its plateau value ($\sim96$\%) for 
$p_{\rm T,\mu}\sim 1.5$ GeV/$c$. 
The offline event selection, the muon reconstruction and identification criteria and the kinematic cuts applied at the
single and dimuon levels are identical to those described in~\cite{Abe14}. In addition, a cut on the transverse distance from the primary vertex of each of the reconstructed muon tracks, weighted with its momentum ($p$DCA), was performed. Tracks with $p$DCA $> 6 \times \sigma_{p{\rm DCA}}$ were rejected. 
The quantity $\sigma_{p{\rm DCA}}$ is the $p$DCA resolution, which is obtained from data, taking into account the resolution on track momentum and slope~\cite{Lop14}. 
Such a track cut reduces the background continuum by a few percent without affecting the resonances.

\begin{figure}[htbp]
\centering
\includegraphics[width=0.48\textwidth,bb=0 0 564 411]{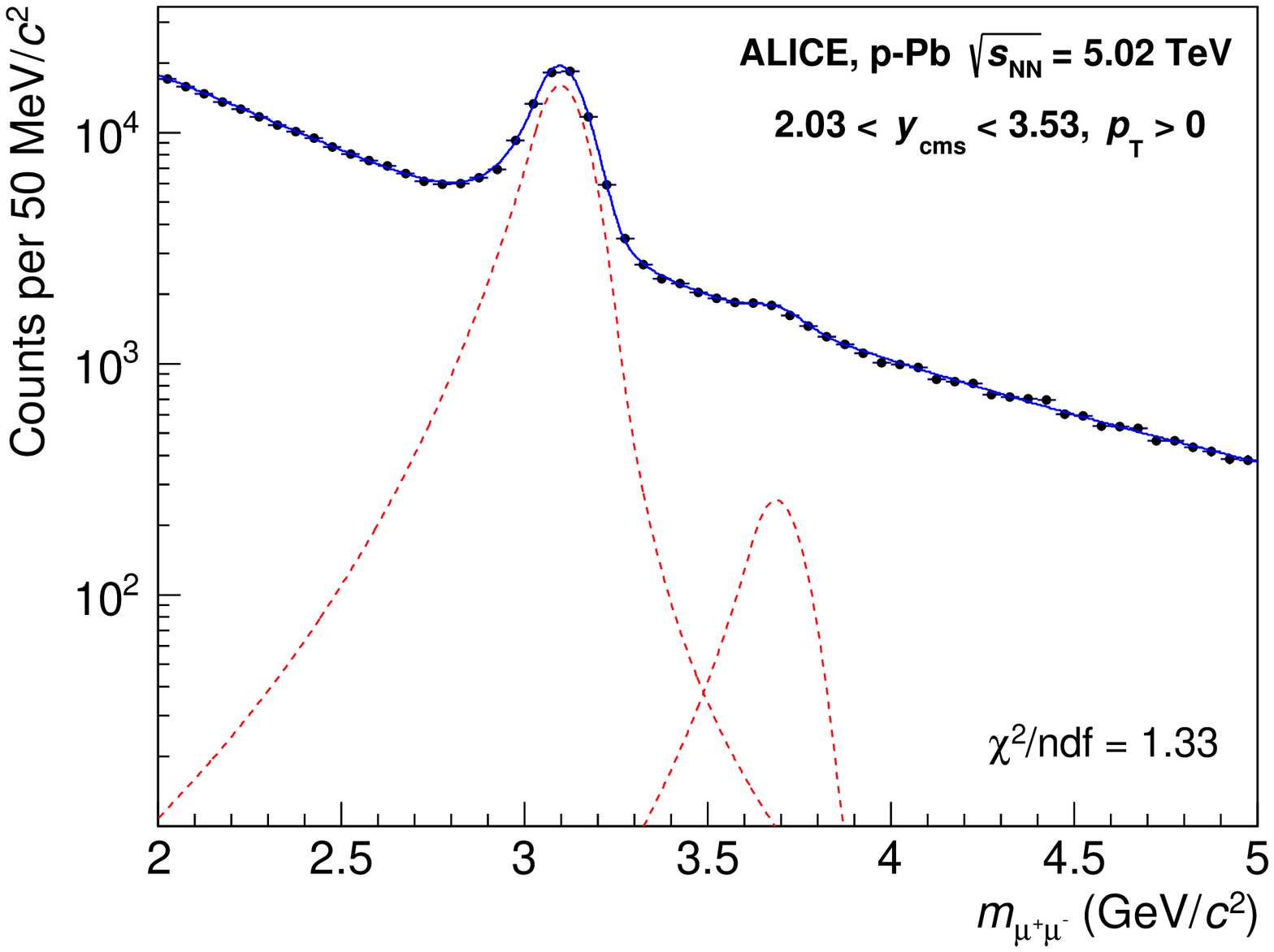}    
\includegraphics[width=0.48\textwidth,bb=0 0 564 411]{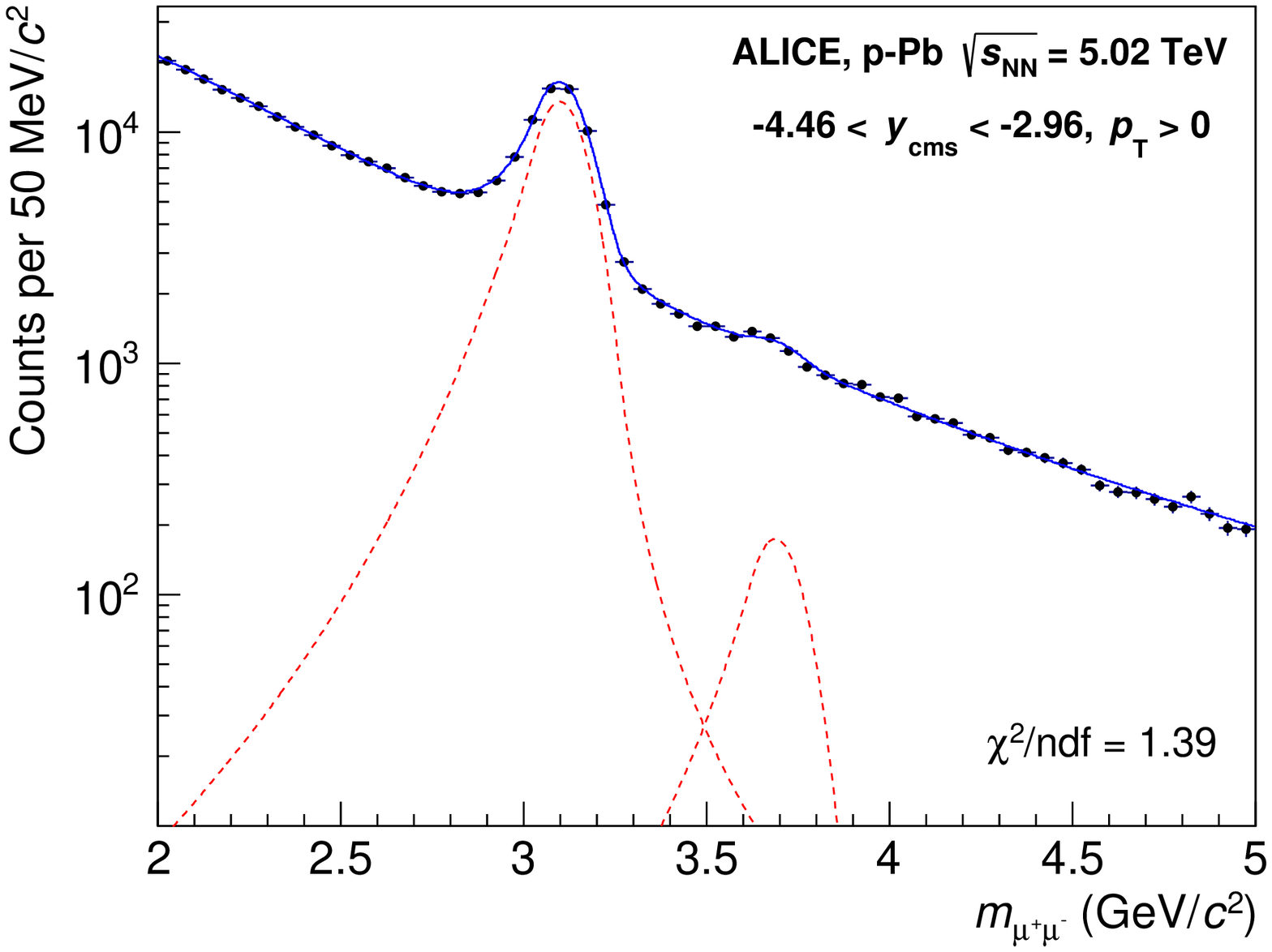} 
\caption{Opposite-sign dimuon invariant mass spectra for the \mbox{p-Pb} (left) and \mbox{Pb-p} (right) data samples, together with the result of a fit. 
For the  fits shown here, Crystal Ball functions (shown as dashed lines) and a variable-width Gaussian have been used for the resonances and the background, 
respectively. The $\chi^{2}$/ndf refers to the goodness of the signal and background combined fit in the displayed mass range.}
\label{fig:1}
\end{figure}

The extraction of the resonance signals is carried out by means of a fit to the
dimuon invariant mass spectrum, as illustrated in Fig.~\ref{fig:1} for the two rapidity ranges under study. The J/$\psi$ and \psip\ line shapes are described either by Crystal Ball (CB) functions~\cite{Gai82}, 
with asymmetric tails on both sides of the peak, or by pseudo-Gaussian functions~\cite{Sha01}. The parameters of the resonance shapes are obtained by means of a Monte-Carlo (MC) simulation. Pure J/$\psi$ and \psip\ signal samples are generated, and then tracked and reconstructed in the experimental setup with the same procedure applied to real data. The choice of the MC kinematic distributions of charmonia is discussed below when introducing the acceptance calculation. Due to the large signal to background ratio (S/B) in the J/$\psi$ mass region and in order to account for small deviations of the mass ($\sim$0.1\%) and width ($\sim$10\%) between MC and data, the corresponding parameters are left free in the fit. For the \psip, due to the less favourable S/B, the mass and widths are constrained by those for the 
J/$\psi$ using the following relations, which involve the 
corresponding MC quantities: 
$m_{\psi(\rm 2S)} = m_{\rm J/\psi}+(m^{\rm MC}_{\psi(\rm 2S)}-m^{\rm MC}_{\rm J/\psi})$ and 
$\sigma_{\psi(\rm 2S)} = \sigma_{\rm J/\psi}\cdot (\sigma^{\rm MC}_{\psi(\rm 2S)}/\sigma^{\rm MC}_{\rm J/\psi})$. 
Alternative values of the \psip\ mass resolution have also been tested, allowing the ratio $(\sigma^{\rm MC}_{\psi(\rm 2S)}/\sigma^{\rm MC}_{\rm J/\psi})$ 
to vary within 10\%~\cite{Lop14}.
Finally, the parameters of the asymmetric tails, which can hardly be constrained by the data, are 
kept fixed to their MC values.  
Additional sets of tails, obtained from the MC, but sampling the $y_{\rm cms}$ and $p_{\rm T}$ phase space, have also been tested. 
The dependence of the extracted J/$\psi$ and $\psi(2{\rm S})$ yields on the variation of the tails and on the \psip\ mass resolution 
is included in the systematic uncertainty on the signal extraction.
The background continuum under the resonances is parameterized by empirical shapes, using a polynomial times
an exponential function or a Gaussian having a width increasing with mass.
In order to assess the systematic uncertainty on signal extraction, fits with various combinations of the signal and background shapes are performed, and the start/end point of the fit range is also varied. 
The  raw \psip\ yields and their statistical uncertainty is finally obtained as the average of the results of the various fits performed, while the systematic uncertainty is calculated as the root-mean-square (RMS) of their distribution. 
This results in $N_{\rm {pPb}}^{\psi(\rm 2S)} = 1069 \pm 130 \pm 102$ and $N_{\rm {Pbp}}^{\psi(\rm 2S)} = 697 \pm 111 \pm 65$, where the first uncertainty is statistical and the second is systematic. The \psip\ mass resolution extracted from the fits is $\sim$70 MeV/$c^2$. As a cross-check, an alternative approach for signal extraction, based on event counting, was also tested. 
More precisely, after fitting the invariant mass distribution and subtracting the background contribution, the number of \psip\ was obtained by integrating the background subtracted spectrum in the region $3.5< m_{\mu\mu} < 3.8$ GeV/$c^2$. 
Corrections, based on the signal fitting functions, were applied to the measured number of counts to account for the fraction of \psip\ outside of the integration region 
($\sim$15\%) and for the number of J/$\psi$ falling inside the \psip\ mass range ($\sim$8\%). 
The results were found to be stable within 1\% with respect to 0.1 GeV/$c^2$ variations of the integration region.
The number of \jpsi\ and \psip\ extracted in this way are also in excellent agreement (i.e., well within the systematic 
uncertainties) with respect to the $N_{\rm {pPb}}^{\psi(\rm 2S)}$ and $N_{\rm {Pbp}}^{\psi(\rm 2S)}$ values quoted above.

The acceptance times efficiency values ($\rm{A}\times \epsilon$) for the \psip\ were evaluated 
using MC simulations in a similar way as detailed in~\cite{Abe14} for the J/$\psi$. The input
$p_{\rm T}$ distributions were obtained from those used for the J/$\psi$~\cite{Abe14}, scaled such that $\langle p_{\rm T}\rangle^{\psi(\rm 2S)}_{\rm pPb, 5.02 TeV} = \langle p_{\rm T}\rangle^{{\rm J}/\psi}_{\rm pPb, 5.02 TeV}\times (\langle p_{\rm T}\rangle^{\psi(\rm 2S)}_{\rm pp, 7 TeV}/\langle p_{\rm T}\rangle^{{\rm J}/\psi}_{\rm pp, 7 TeV})$, and using the $\sqrt{s}=7$ TeV pp values from LHCb~\cite{Aai13,Aai12} obtained in the slightly larger range 
$2 < y_{\rm cms} < 4.5$. The input $y$ distributions were obtained from those
used for the J/$\psi$ assuming a scaling of the widths with $y_{\rm
max}^{\psi(\rm 2S)}/y_{\rm max}^{{\rm J}/\psi}$, where $y_{\rm max}^i =
\log(\sqrt{s}/m_i)$ is the maximum rapidity for the resonance $i$ at the
$\sqrt{s}$ value under study. An unpolarized distribution for the \psip\ was assumed,
according to the results obtained in pp collisions at $\sqrt{s} = 7$ TeV by the CMS and LHCb experiments~\cite{Cha13,Aai14}. The systematic uncertainty 
for the \psip\ acceptance was calculated as the maximum spread of the values obtained by assuming as alternative input 
distributions those used for the J/$\psi$ itself and amounts to 1.8\% (2.5\%) for \mbox{p-Pb} (\mbox{Pb-p}).

The efficiency of the tracking and trigger detectors of the MS 
was taken into account in the MC simulations by means of a map of
dead channels (tracking) and by building efficiency tables for the detector
elements (trigger). The evolution of the detector performance throughout the data taking was followed in the MC, by generating a number of events which is proportional to the run-by-run number of dimuon triggers, in order to properly weight the detector conditions over the entire data taking. The
systematic uncertainties on the efficiencies were obtained with algorithms based
on real data, with the same procedure adopted in~\cite{Abe14}, and
they are identical for J/$\psi$ and \psip. A small uncertainty
related to the efficiency of the matching between tracking and triggering
information  was also included~\cite{Abe14}.

The $p_{\rm T}$-integrated $\rm{A}\times \epsilon$ values for \psip\ production, obtained with this
procedure, are $0.270 \pm 0.014$ (\mbox{p-Pb}) and $0.184 \pm 0.013$ (\mbox{Pb-p}),  where 
the lower value for \mbox{Pb-p} is mainly due to a smaller detector efficiency in the corresponding data 
taking period, related to a worse detector performance. The quoted uncertainties are systematic and are obtained as the quadratic sum of the uncertainties on MC input, tracking, triggering and matching efficiencies. The statistical uncertainties are negligible.

The cross section times the branching ratio ${\rm B.R.}(\psi(\rm 2S)\rightarrow\mu\mu)$ for inclusive \psip\ production in \mbox{p-Pb} collisions (and similarly for \mbox{Pb-p}) is: 
\begin{equation}
{\rm B.R.}_{\psi(\rm 2S)\rightarrow\mu^+\mu^-}\cdot\sigma^{\rm \psi(\rm 2S)}_{\rm pPb}=\frac{{\it N}^{\rm cor}_{\rm \psi(\rm 2S)\rightarrow\mu\mu}}
{{\it L}_{\rm int}^{\rm pPb}}
\end{equation}

{\noindent where ${N^{\rm cor}_{\psi(\rm 2S)\rightarrow\mu\mu}}$ is the number of \psip\ corrected for 
$\rm{A}\times \epsilon$, 
and ${\it L}_{\rm int}^{\rm pPb}$ is the integrated luminosity, calculated as $N_{\rm MB}/\sigma^{\rm MB}_{\rm pPb}$.  
$N_{\rm MB}$ is the number of MB events, obtained as the number of dimuon triggers divided by the
probability of having a triggered dimuon in a MB event. The $N_{\rm MB}$ numerical values 
and uncertainties are the same as those quoted in~\cite{Abe14}.  
The cross sections for the occurrence of the MB condition, $\sigma^{\rm MB}_{\rm pPb}$, are measured in a vdM scan~\cite{Gag13} to be 2.09 $\pm$ 0.07 b for the \mbox{p-Pb} configuration and
2.12 $\pm$ 0.07 b for the \mbox{Pb-p} one. The luminosity is also
independently determined by means of a second luminosity signal,
as described in~\cite{Gag13}. The two measurements differ by at most
1\% throughout the whole data-taking period and such a value is quadratically
added to the luminosity uncertainty.
The \psip\ cross section values are:}

\begin{eqnarray*}
{\rm B.R.}\cdot\sigma^{\psi(\rm 2S)}_{\rm pPb} (2.03<y_{\rm cms}<3.53) = 0.791 \pm 0.096 {\rm (stat.)} \pm 0.091 {\rm (syst.
uncorr.)} \pm 0.013 {\rm (syst. corr.)}\,\, \mu \rm{b} \\
{\rm B.R.}\cdot\sigma^{\psi(\rm 2S)}_{\rm Pbp}(-4.46<y_{\rm cms}<-2.96) = 0.653 \pm 0.104 {\rm (stat.)} \pm 0.080 {\rm (syst.
uncorr.)} \pm 0.010 {\rm (syst. corr.)}\,\, \mu \rm{b} 
\end{eqnarray*}

The systematic uncertainties for the \psip\ cross section measurement are obtained as the quadratic sum of the various contributions listed in Table~\ref{tab:1}. The splitting between uncorrelated and correlated sources is also summarized there. The corresponding values for the J/$\psi$ can be found in~\cite{Abe14}.

\begin{table}
\centering
\begin{tabular}{c|c|c}
\hline
 & B.R.$\cdot\sigma^{\psi(\rm 2S)}_{\rm pPb}$ & B.R.$\cdot\sigma^{\psi(\rm 2S)}_{\rm Pbp}$ \\ \hline
Tracking efficiency & 4 & 6 \\
Trigger efficiency & 2.8 (2 $-$ 3.5) & 3.2 (2 $-$ 3.5) \\
Signal extraction & 9.5 (8 $-$ 11.9) & 9.3 (8.6 $-$ 12.7) \\
MC input & 1.8 (1.5 $-$ 1.5) & 2.5 (1.5 $-$ 1.7) \\
Matching efficiency & 1 & 1 \\ 
$L_{\rm int}$(uncorr.) & 3.4 & 3.1\\
$L_{\rm int}$(corr.) & 1.6 & 1.6\\
\end{tabular}
\caption{\label{tab:1} Systematic uncertainties (in percent) affecting the measurement of inclusive
\psip\ cross sections. 
The $L_{\rm int}$ uncertainties are splitted in two components, respectively uncorrelated and correlated between \mbox{p-Pb} and \mbox{Pb-p}, as detailed in~\cite{Gag13}.
All the other uncertainties are uncorrelated between forward and backward rapidity. 
Uncertainties refer to $p_{\rm T}$-integrated quantities and, where they depend on $p_{\rm T}$, the corresponding maximum and minimum values are also quoted.
The efficiency-related uncertainties refer to muon pairs.}
\end{table}

The study of the cross section ratio between \psip\ and J/$\psi$, and the comparison of this ratio between different systems, offers a powerful 
tool to investigate nuclear effects on charmonium production. In addition, several systematic uncertainties cancel, or are significantly reduced, when studying such ratios. In particular, in the present analysis, the tracking, trigger and matching efficiencies, as well as the normalization-related quantities, cancel out.  For the MC input, the fraction of the uncertainty related to the choice of the J/$\psi$ kinematical distribution~\cite{Abe14} cancels in the cross section ratios, 
and the remaining 1\% (2\%) uncertainty for \mbox{p-Pb} (\mbox{Pb-p}) is assigned to this source. 
Finally, the uncertainty on signal extraction is considered as uncorrelated between J/$\psi$ and \psip, and its value for the cross section ratios 
amounts to 10\% for both \mbox{p-Pb} and \mbox{Pb-p}. 
The resulting values are:

\begin{eqnarray*}
\frac{{\rm B.R.}_{\psi(\rm 2S)\rightarrow\mu^+\mu^-}\sigma^{\psi(\rm 2S)}}{{\rm B.R.}_{{\rm J}/\psi\rightarrow\mu^+\mu^-}\sigma^{{\rm J}/\psi}}(2.03<y_{\rm cms}<3.53) = 
 0.0154 \pm 0.0019 {\rm (stat.)} \pm 0.0015 {\rm (syst.)}  \\
\frac{{\rm B.R.}_{\psi(\rm 2S)\rightarrow\mu^+\mu^-}\sigma^{\psi(\rm 2S)}}{{\rm B.R.}_{{\rm J}/\psi\rightarrow\mu^+\mu^-}\sigma^{{\rm J}/\psi}}(-4.46<y_{\rm cms}<-2.96) = 
 0.0116 \pm 0.0018 {\rm (stat.)} \pm 0.0011 {\rm (syst.)} 
\end{eqnarray*}

\begin{figure}[htbp]
\centering
\includegraphics[width=0.7\textwidth,bb=0 0 564 411]{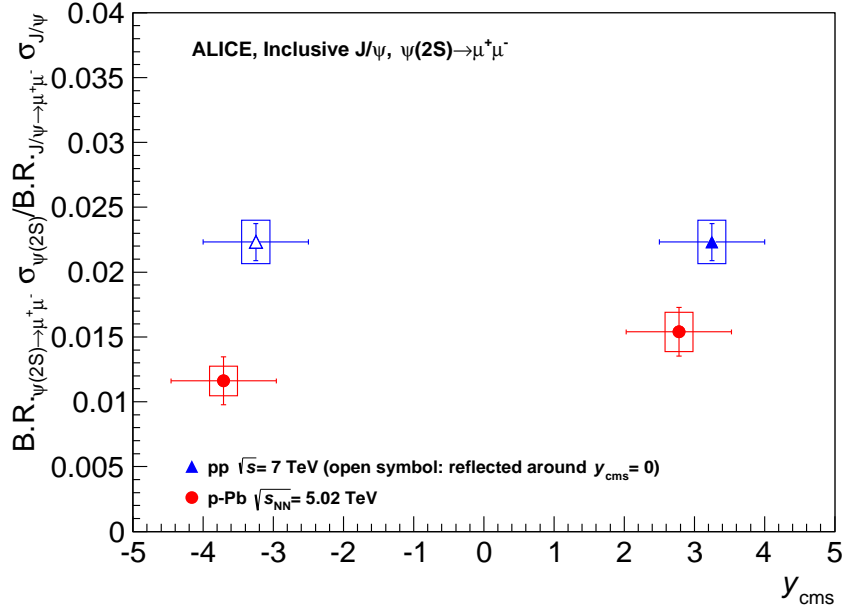}
\caption{The cross section ratios ${\rm B.R.}_{\psi(\rm 2S)\rightarrow\mu^+\mu^-}\sigma^{\psi(\rm 2S)}/{\rm B.R.}_{{\rm J}/\psi\rightarrow\mu^+\mu^-}\sigma^{{\rm J}/\psi}$ for \mbox{p-Pb} and \mbox{Pb-p} collisions, 
compared with the corresponding pp results at $\sqrt{s}=7$ TeV~\cite{Lop14}. The horizontal bars correspond to the width of the rapidity regions under study.
The vertical error bars represent statistical uncertainties, the boxes correspond to systematic uncertainties.}
\label{fig:2}
\end{figure}

In Fig.~\ref{fig:2} we compare these ratios with the corresponding ALICE results for pp collisions~\cite{Lop14}, obtained in  slightly different centre of mass energy and rapidity regions, $\sqrt{s}$ = 7 TeV, $2.5<|y|<4$, as no LHC pp results are available in the same kinematic conditions of proton-nucleus collisions. 
The pp ratios are significantly higher than those for \mbox{p-Pb} and \mbox{Pb-p}, which are compatible within
uncertainties.

The double ratio 
$[\sigma_{\psi(\rm 2S)}/\sigma_{\rm J/\psi}]_{\rm pPb}/[\sigma_{\psi(\rm 2S)}/\sigma_{\rm J/\psi}]_{\rm pp}$ 
is a useful quantity to directly compare the relative suppression of the two states between various 
experiments.
For this analysis, since the collision energy and the $y$-coverage of the \mbox{p-Pb} (\mbox{Pb-p}) and pp measurements
are different, we have estimated the possible dependence of the $\sigma^{\psi(\rm 2S)}/\sigma^{{\rm J}/\psi}$ vs
$\sqrt{s}$ and $y$ in pp collisions. We start from the empirical observation that this ratio is very similar at collider
energies over a rather broad range of $y$ and $\sqrt{s}$. In particular, from the LHCb data ($\sqrt{s}$ = 7 TeV,
$2<y<4.5$)~\cite{Aai13, Aai12} one gets 2.11\% for the inclusive ratio integrated over $p_{\rm T}$, while the
corresponding value from CDF data (p${\rm \overline p}$ at $\sqrt{s}$ = 1.96 TeV, $|y|<0.6$)~\cite{Aal09} is 2.05\%, i.e., only 3\% smaller  
(the latter quantity was obtained by extrapolating the CDF \psip\ measurement to $p_{\rm T}=0$ with 
the phenomenological function $f(p_{\rm T})= (p_{\rm T})/[1+(p_{\rm T}/a)^2]^b$)~\cite{Ada07}. The LHCb result can be extrapolated to central rapidity at $\sqrt{s}$ = 7 TeV, assuming a Gaussian $y$-distribution for both resonances, with the width of the J/$\psi$ distribution tuned directly on data~\cite{Aai13} and that for \psip\ obtained from the former assuming a scaling of the widths with $y^{\psi(\rm 2S)}_{\rm max}/y^{{\rm J}/\psi}_{\rm max}$. The effect of this rescaling is small, leading to a 3\% increase of the ratio. The central-rapidity ratio $\sigma_{\psi(\rm 2S)}/\sigma_{\rm J/\psi}$ at $\sqrt{s}=5.02$ TeV is then obtained by means of an interpolation between the CDF and LHCb-rescaled values, assuming a linear dependence of the ratio vs $\sqrt{s}$. Finally, one can extrapolate the ratio to the \mbox{p-Pb} and \mbox{Pb-p} rapidity ranges by using for the J/$\psi$ the  Gaussian shape obtained with the interpolation procedure described in~\cite{ALPN} and for the $\psi(\rm 2S)$ the corresponding shape 
scaled with 
$y^{\psi(\rm 2S)}_{\rm max}/y^{{\rm J}/\psi}_{\rm max}$. 
The difference between the measured value of $\sigma_{\psi(\rm 2S)}/\sigma_{\rm J/\psi}$ for $\sqrt{s}$ = 7 TeV, $2<y_{\rm cms}<4.5$ and the results of the interpolation 
procedure to $\sqrt{s} = 5.02$ TeV, $2.03<y_{\rm cms}<3.53$ ($-4.46<y_{\rm cms}<-2.96$) is -1.6\% (-3.7\%). 
When calculating the double ratio $[\sigma_{\psi(\rm 2S)}/\sigma_{\rm J/\psi}]_{\rm pPb}/[\sigma_{\psi(\rm 2S)}/\sigma_{\rm J/\psi}]_{\rm pp}$, we choose to use for pp the measured value at $\sqrt{s}$ = 7 TeV, $2.5<y_{\rm cms}<4$~\cite{Lop14} (rather than the interpolated one at $\sqrt{s} = 5.02$ TeV) and to 
 include a 8\% systematic uncertainty on this quantity, i.e., about twice the maximum difference between the measured 
 values of the ratio in pp and the results of the interpolation procedure. A similar uncertainty would be obtained
 using as an input for the calculation, instead of the LHCb data, the more recent pp result from ALICE on $\sigma_{\psi(\rm 2S)}/\sigma_{\rm
 J/\psi}$~\cite{Lop14}.

\begin{figure}[htbp]
\centering
\includegraphics[width=0.7\textwidth,bb=0 0 564 411]{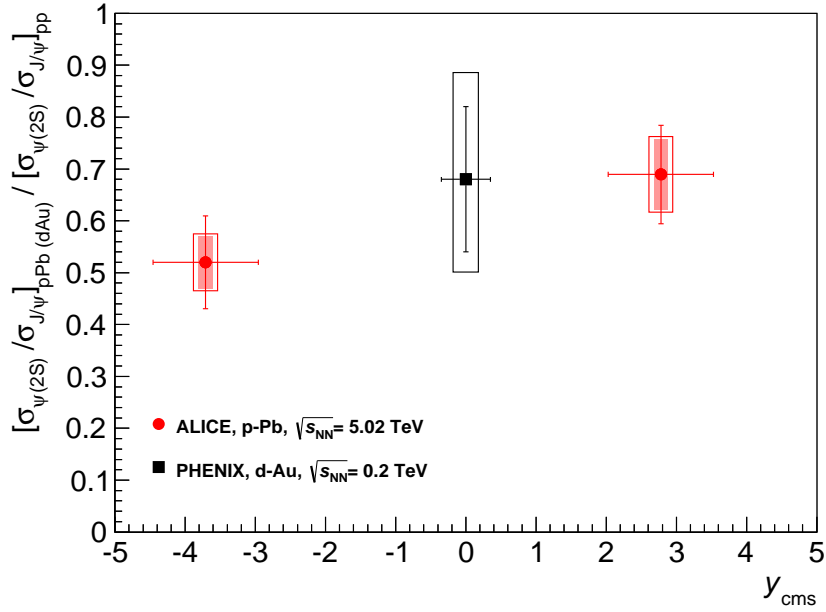}
\caption{Double ratios $[\sigma_{\psi(\rm 2S)}/\sigma_{\rm J/\psi}]_{\rm pPb}/[\sigma_{\psi(\rm 2S)}/\sigma_{\rm J/\psi}]_{\rm pp}$ for \mbox{p-Pb} and \mbox{Pb-p} collisions, 
compared to the corresponding PHENIX result at $\sqrt{s_{\rm NN}}$~=~200 GeV~\cite{Ada13}. The horizontal bars correspond to the width of the rapidity regions under study. For ALICE, the vertical error bars correspond to statistical uncertainties, the boxes to uncorrelated systematic uncertainties, and the shaded areas to correlated uncertainties. For PHENIX, the various sources of systematic uncertainties were combined in quadrature.}
\label{fig:3}
\end{figure}

The values of the double ratio are shown in Fig.~\ref{fig:3}, where they are also compared with the corresponding results obtained by the PHENIX experiment at $\sqrt{s_{\rm NN}}$~=~200 GeV, for $|y|<0.35$~\cite{Ada13}. When forming the double ratio, the systematic uncertainties on the pp ratio, including the 8\% contribution described in the previous paragraph, are considered as correlated between forward and backward rapidity, while the other systematic uncertainties are treated as uncorrelated. The dominating contributions to the systematic uncertainty come from the signal extraction and from the interpolation procedure used for the pp cross section.
The ALICE results show that, compared to pp, the \psip\ is more suppressed than the J/$\psi$ to a 2.3$\sigma$ (4.1$\sigma$) level in \mbox{p-Pb} (\mbox{Pb-p}). 
The PHENIX result shows a similar feature, at a 1.3$\sigma$ level. 

\begin{figure}[htbp]
\centering
\includegraphics[width=0.7\textwidth,bb=0 0 564 411]{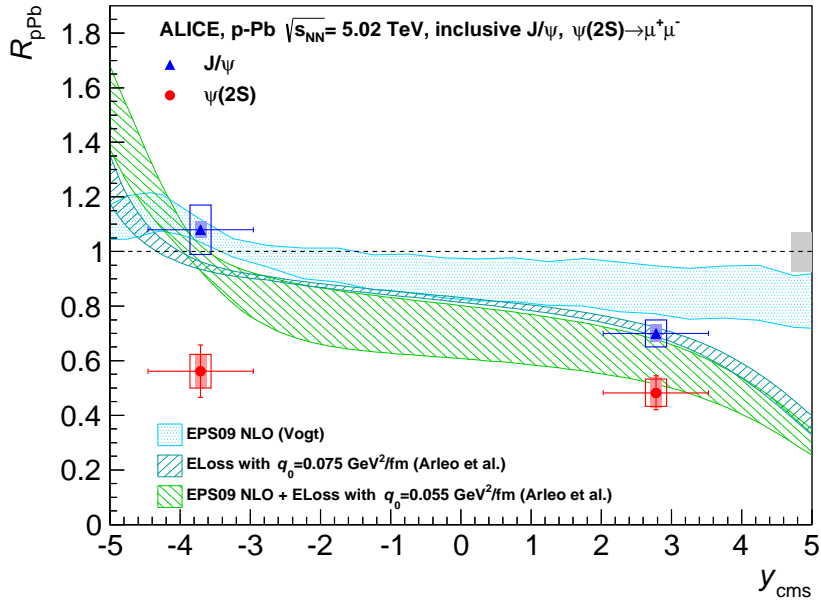}
\caption{The nuclear modification factor for $\psi(\rm 2S)$, compared to the corresponding quantity for J/$\psi$~\cite{Abe14}. The horizontal bars correspond to the width of the rapidity regions under study.
The vertical error bars correspond to statistical uncertainties, the boxes to uncorrelated systematic uncertainties, and the shaded areas to partially correlated uncertainties. 
The filled box on the right, centered on $R_{\rm pPb}=1$, shows uncertainties that are fully correlated between J/$\psi$ and $\psi(\rm 2S)$. Model calculations tuned on J/$\psi$, and including nuclear shadowing~\cite{Alb13} and coherent energy loss~\cite{Arl13} are also shown. 
The corresponding calculations for \psip\ produce identical values for the coherent energy loss mechanisms and a 2-3\% larger result for nuclear shadowing and therefore are not shown.}
\label{fig:4}
\end{figure}

The suppression of charmonium states with respect to the corresponding pp yield can be quantified using the nuclear modification factor. For \psip, $R^{\psi(\rm 2S)}_{\rm pPb}$ is obtained by combining $R^{{\rm J}/\psi}_{\rm pPb}$~\cite{Abe14} with the double ratio evaluated above:

\begin{equation}
R^{\psi(\rm 2S)}_{\rm pPb}=R^{{\rm J}/\psi}_{\rm pPb}\cdot \frac{\sigma^{\psi(\rm 2S)}_{\rm pPb}}
{\sigma^{{\rm J}/\psi}_{\rm pPb}}\cdot 
\frac{\sigma^{{\rm J}/\psi}_{\rm pp}}{\sigma^{\psi(\rm 2S)}_{\rm pp}}
\label{eq:1}
\end{equation}

In Fig.~\ref{fig:4}, $R^{\psi(\rm 2S)}_{\rm pPb}$ is shown and compared with
$R^{{\rm J}/\psi}_{\rm pPb}$. For the double ratios, the difference in the
$\sqrt{s}$ and $y$ domains between \mbox{p-Pb} and pp is taken into account by the inclusion of the 8\% systematic 
uncertainty described above. The other quoted uncertainties combine those from $R^{{\rm J}/\psi}_{\rm pPb}$~\cite{Abe14} 
with those for the double ratio, avoiding a double counting of the J/$\psi$ related uncertainties. 
Figure~\ref{fig:4} indicates that the \psip\ suppression is much stronger than for the J/$\psi$ and reaches a factor $\sim$2 with respect to pp. 
The results are compared with theoretical calculations including either nuclear shadowing only~\cite{Alb13} or coherent energy loss, with or without a shadowing contribution~\cite{Arl13}.
For the former mechanism, the values correspond to calculations performed for the J/$\psi$. 
However, due to
the relatively similar kinematic distributions of gluons that produce the $c\overline c$ pair which will then hadronize 
to a J/$\psi$ or a \psip, the shadowing effects are expected to be the same,  within 2-3\%~\cite{Fer14}, for the two 
charmonium states. No sensitivity to the final quantum numbers of the charmonium state is expected for coherent energy 
loss, implying that the calculations shown in Fig.~\ref{fig:4} are valid for both resonances. 
As a consequence, all three models would predict an almost identical suppression for the \psip\ and the \jpsi\ over the
full rapidity range, with negligible theoretical uncertainties. This prediction is in strong disagreement with our data 
and clearly indicates that other mechanisms must be invoked in 
order to describe the \psip\ suppression in proton-nucleus collisions.

The break-up cross section of the final state resonance due to interactions with CNM is expected to depend on the binding energy of the charmonium and such a mechanism would be a natural explanation for the larger suppression of \psip. However, this process becomes relevant only if the charmonium formation time $\tau_{\rm f}$ is smaller than the time $\tau_{\rm c}$ spent by the $c\overline c$ pair inside the nucleus. One can evaluate the average proper time $\tau_{\rm c}$ spent in CNM as $\tau_{\rm c}=\langle L\rangle/(\beta_{\rm z}\gamma)$~\cite{McG13}, where $\langle L\rangle$ is the average length of nuclear matter crossed by the pair, 
which can be calculated in the framework of the Glauber model~\cite{Mil07}, $\beta_{\rm z}=\tanh y_{c\overline c}^{\rm rest}$ is the velocity of the $c\overline c$ along the beam direction in the nucleus rest frame, and $\gamma=E_{c\overline c}/m_{c\overline c}$. For $c\overline c$ pairs in the charmonium mass range emitted at $p_{\rm T}=0$ in the forward 
acceptance, one gets 
$\tau_{\rm c}\sim 10^{-4}$ fm/$c$, while the corresponding value at backward rapidity is $\tau_{\rm c}\sim 7\cdot
10^{-2}$ fm/$c$. Estimates for the formation time $\tau_{\rm f}$ range between 0.05 and 0.15
fm/$c$~\cite{Arl00,McG13}. In this situation, no break-up effects depending on the final charmonium state
should be expected at forward rapidity, and even for backward production one has at most $\tau_{\rm f} \sim
\tau_{\rm c}$ which would hardly accomodate the strong difference observed between \psip\ and 
J/$\psi$ suppression. As a consequence, other final state effects should be considered, 
including the interaction of the $c\overline c$ pair with the final state hadronic system created in the 
proton-nucleus collision.

\begin{figure}[h!]
\centering
\includegraphics[width=1\textwidth,bb=0 0 564 275]{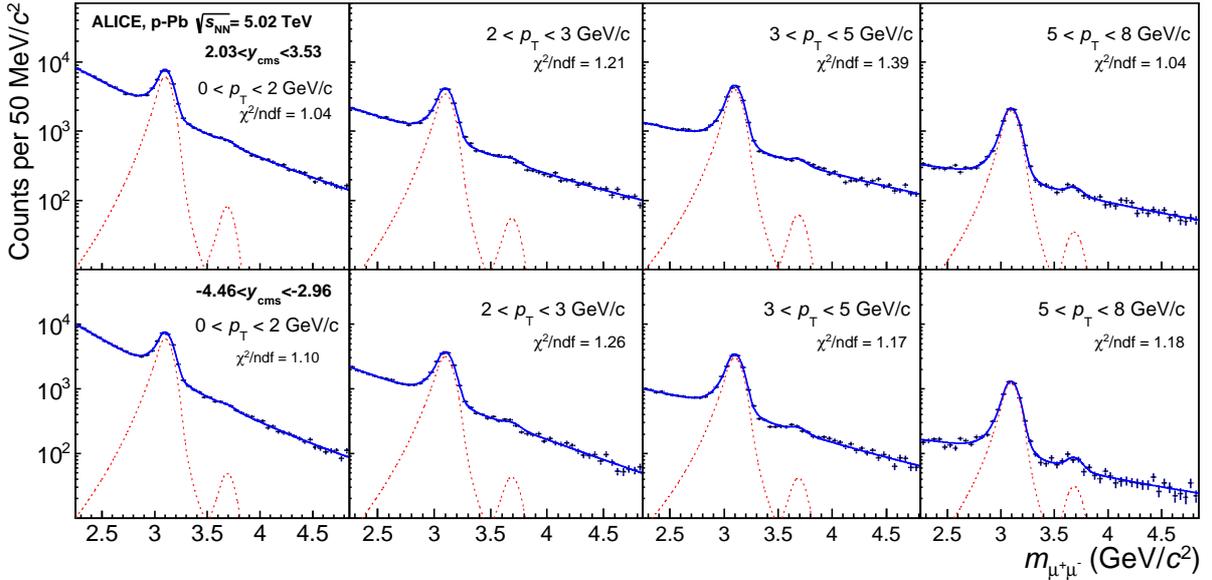}
\caption{Opposite-sign dimuon invariant mass spectra, in bins of transverse momentum, for the \mbox{p-Pb} and \mbox{Pb-p} data samples. 
For the fits shown here, Crystal Ball functions (shown as dashed lines) and a variable-width Gaussian have been used for the resonances and the background, respectively.
The $\chi^{2}$/ndf refers to the goodness of the signal and background combined fit in the displayed mass range.}
\label{fig:added}
\end{figure}

\begin{figure}[h!]
\centering
\includegraphics[width=0.7\textwidth,bb=0 0 564 411]{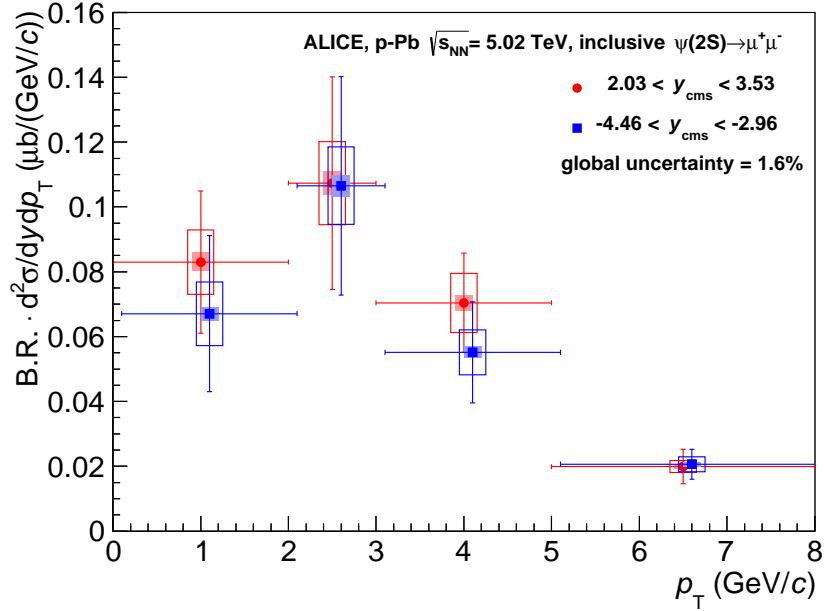}
\caption{The \psip\ differential cross sections B.R.$\cdot{\rm d}^2\sigma/{\rm d}y{\rm d}p_{\rm T}$ for \mbox{p-Pb}
and \mbox{Pb-p} collisions. The horizontal bars correspond to the width of the transverse momentum bins. The
vertical error bars correspond to the statistical uncertainties, the boxes to uncorrelated systematic
uncertainties and the shaded areas to $p_{\rm T}$-correlated uncertainties. A global 1.6\% uncertainty applies to both \mbox{p-Pb} and \mbox{Pb-p} results.
The points corresponding to negative $y$ are slightly shifted in $p_{\rm T}$ to improve visibility.}
\label{fig:5}
\end{figure}

The sizeable \psip\ statistics collected in proton-nucleus collisions allows for a differential study of the
various observables as a function of $p_{\rm T}$, in the range $0<p_{\rm T}<8$ GeV/$c$. We have chosen a
transverse momentum binning which leads to similar relative statistical uncertainties in each bin over the
$p_{\rm T}$ range covered. The analysis is carried out with the same procedure adopted for the integrated
data samples. In particular, the systematic uncertainties are evaluated differentially in $p_{\rm T}$, and
their range is also reported in Table~\ref{tab:1}. In Fig.~\ref{fig:added} the invariant mass spectra for the various $p_{\rm T}$ bins are shown, together with the result of the fits. In Fig.~\ref{fig:5} the differential cross sections at
forward and backward rapidity are presented. The systematic uncertainties on signal extraction, MC input and
efficiencies are considered as bin-to-bin uncorrelated. 
The $L_{\rm int}$ uncertainties are correlated between the various $p_{\rm T}$ bins and partially correlated between \mbox{p-Pb} and \mbox{Pb-p}. 

\begin{figure}[htbp]
\centering
\includegraphics[width=0.7\textwidth,bb=0 0 564 411]{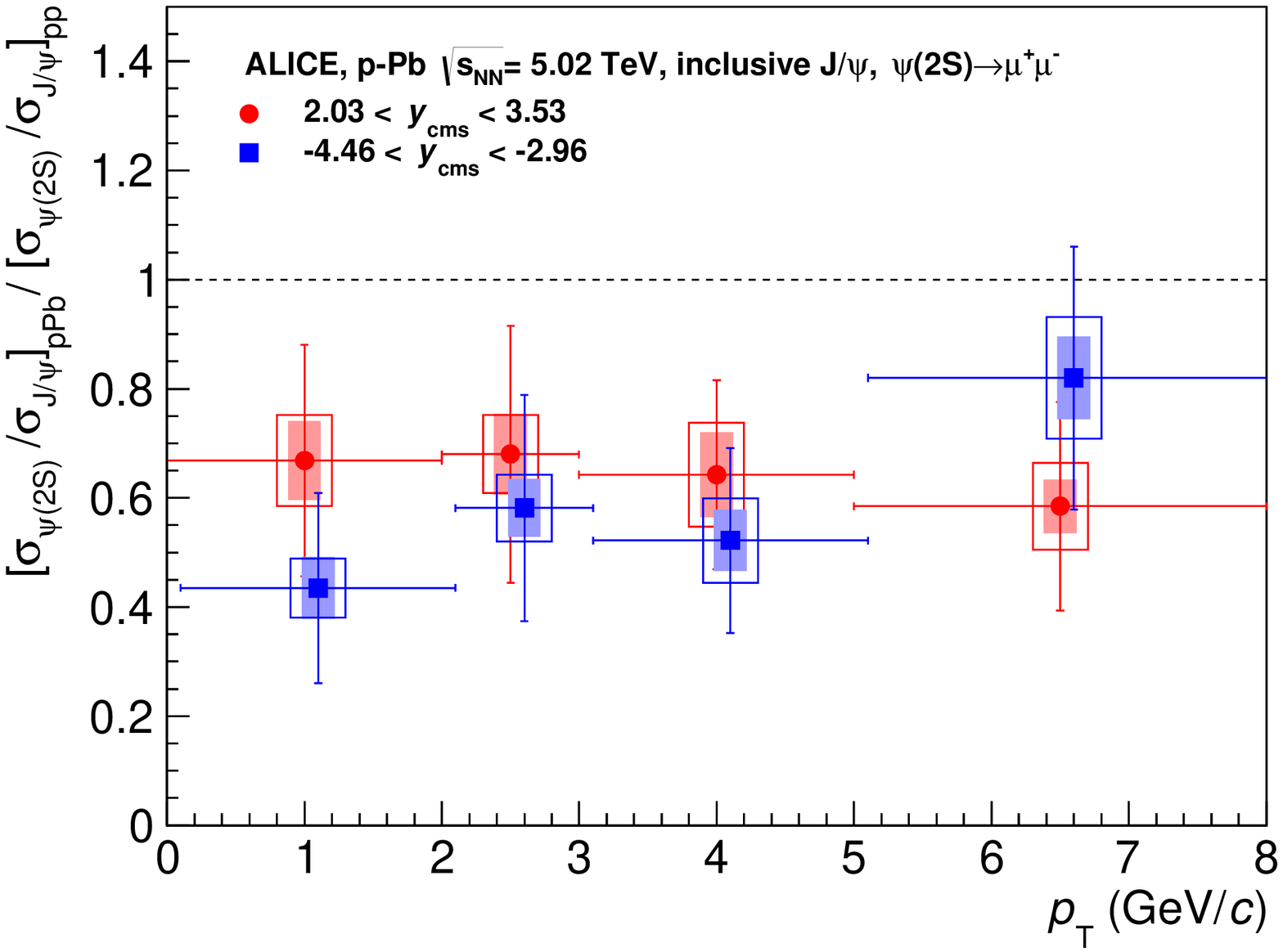}
\caption{The double ratio $[\sigma_{\psi(\rm 2S)}/\sigma_{\rm J/\psi}]_{\rm pPb}/[\sigma_{\psi(\rm
2S)}/\sigma_{\rm J/\psi}]_{\rm pp}$ for \mbox{p-Pb} and \mbox{Pb-p} collisions, as a function of $p_{\rm
T}$.  The horizontal bars correspond to the width of the transverse momentum bins. The vertical error bars
correspond to the statistical uncertainties, the boxes to uncorrelated systematic uncertainties and the
shaded areas to correlated uncertainties. The points corresponding to negative $y$ are slightly shifted 
in $p_{\rm T}$ to improve visibility.}
\label{fig:6}
\end{figure}

In Fig.~\ref{fig:6} we present the $p_{\rm T}$ dependence of the double ratio $[\sigma_{\psi(\rm 2S)}/\sigma_{\rm J/\psi}]_{\rm
pPb}/[\sigma_{\psi(\rm 2S)}/\sigma_{\rm J/\psi}]_{\rm pp}$, with the \mbox{p-Pb} J/$\psi$ cross sections taken from~\cite{Abe14}
and the pp values from~\cite{Lop14}. As for the integrated double ratio, the systematic uncertainties related to efficiencies and
to normalizations cancel out for both proton-nucleus and pp, while the uncertainties on signal extraction and Monte-Carlo input
are considered as uncorrelated. The 8\% uncertainty related to the $\sqrt{s}$ and $y$ mismatch between the two systems is
correlated as a function of $p_{\rm T}$, while the uncertainties on the ratio in \mbox{pp} collisions are correlated, for each $p_{\rm T}$ bin, between forward and backward rapidity. 

\begin{figure}[htbp]
\centering
\includegraphics*[width=0.49\textwidth,bb=20 0 564 411]{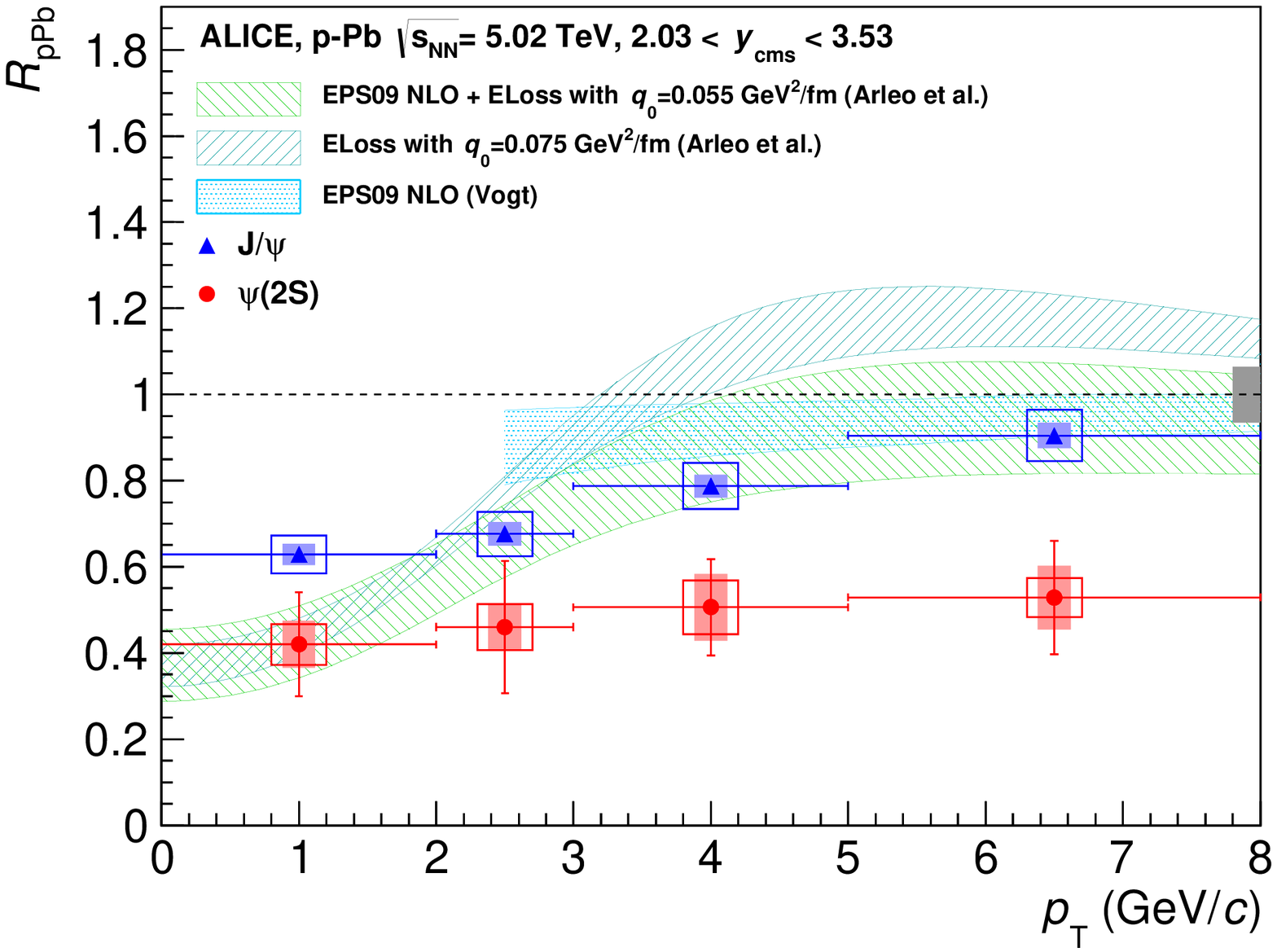}
\includegraphics*[width=0.49\textwidth,bb=20 0 564 411]{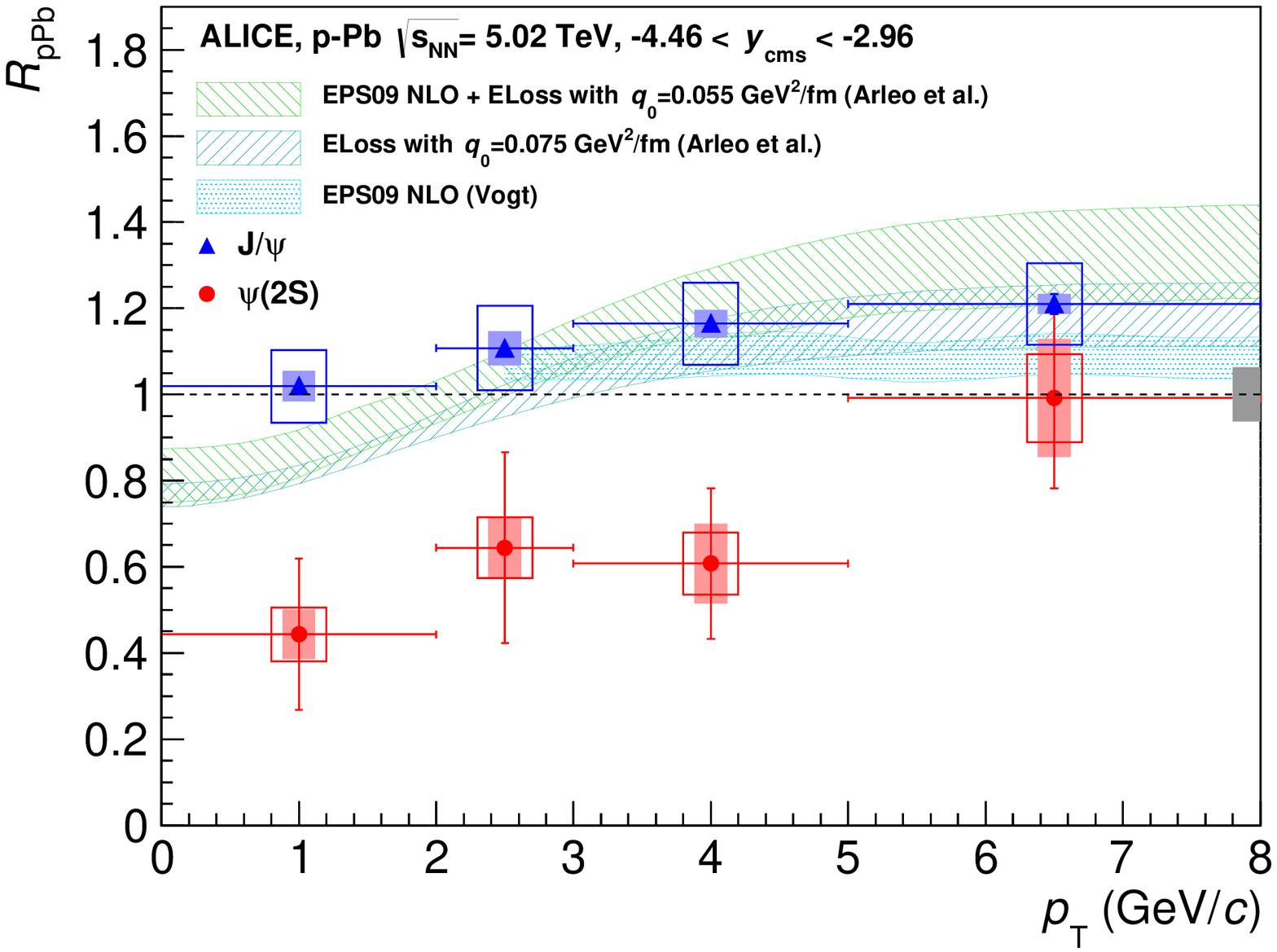}
\caption{The nuclear modification factor for $\psi(\rm 2S)$, compared to the corresponding quantity for 
J/$\psi$~\cite{Arn14}, as a function of $p_{\rm T}$. Plots correspond to \mbox{p-Pb} (left) and \mbox{Pb-p}
(right) collisions.
The horizontal bars correspond to the width of the transverse momentum bins.
The vertical error bars correspond to statistical uncertainties, the boxes to uncorrelated systematic uncertainties, and the shaded areas to partially 
correlated uncertainties. The filled box on the right, centered at $R_{\rm pPb}=1$, shows uncertainties that are fully correlated between J/$\psi$ and $\psi(\rm 2S)$. 
For details on model comparisons, see the caption of Fig.~\ref{fig:4}.}
\label{fig:7}
\end{figure}

Finally, in Fig.~\ref{fig:7} the $p_{\rm T}$ dependence of the \psip\ nuclear modification factor, calculated using Eq.~\ref{eq:1}, is presented and 
compared with the corresponding result for J/$\psi$~\cite{Arn14}. The uncertainties are obtained with the procedure 
used in Fig.~\ref{fig:4}, and the results are compared to the same models quoted there.

Within uncertainties, no $p_{\rm T}$ dependence of the double ratio can be seen, and consequently as a function of transverse momentum $R^{\psi(\rm 2S)}_{\rm pPb}$ has qualitatively a similar
shape as that exhibited by $R^{{\rm J}/\psi}_{\rm pPb}$, but systematically characterized by smaller values. 
Theoretical models, which in this case also yield the same prediction for J/$\psi$ and \psip,
are in fair agreement with J/$\psi$ results, but clearly overestimate the \psip\ nuclear modification
factor values.

It is interesting to note that different values of transverse momentum for the resonances correspond to different $\tau_{\rm c}$, with the crossing times decreasing with increasing $p_{\rm T}$. In particular, for backward production, $\tau_{\rm c}$ varies by about a factor 2, between $\sim$0.07 (at $p_{\rm T}=0$) and $\sim$0.03 fm/$c$ (at $p_{\rm T}=8$ GeV/$c$). As a consequence, a larger fraction of $c\overline c$ pairs may form the final resonance state inside CNM at low $p_{\rm T}$, and one might expect smaller values of the double ratio in that transverse momentum region due to the weaker binding energy of \psip. Although the results shown in Fig.~\ref{fig:6} could be suggestive of such a trend, no firm conclusion can be reached due to the current experimental  uncertainties. 

In summary, we have presented results on inclusive \psip\ production in proton-nucleus collisions at the
LHC. Measurements were performed with the ALICE Muon Spectrometer in the p-going ($2.03<y_{\rm cms}<3.53$)
and Pb-going ($-4.46<y_{\rm cms}<-2.96$) directions, and the production cross sections, the double ratios
with respect to the J/$\psi$ in \mbox{p-Pb} and pp and the nuclear modification factors were estimated. The
results show that \psip\ is significantly more suppressed than J/$\psi$ in both rapidity regions, and that
no $p_{\rm T}$ dependence of this effect is found within uncertainties. This observation implies that initial state nuclear effects alone cannot account for the modification of the \psip\ yields, as also confirmed by the poor agreement of the \psip\ $R_{\rm pPb}$ with models based on shadowing and/or energy loss.
Final state effects, such as the pair break-up by interactions with cold nuclear matter, might in 
principle lead to the observed effect, but the extremely short crossing times for the $c\overline c$ pair, in particular 
at forward rapidity, make such an explanation unlikely. 
Consequently, other final state effects should be considered, including the interaction of the $c\overline c$ pair with 
the final state hadronic system created in the proton-nucleus collision. 

%
\newenvironment{acknowledgement}{\relax}{\relax}
\begin{acknowledgement}
\section*{Acknowledgements}
The ALICE collaboration would like to thank all its engineers and technicians for their invaluable contributions to the construction of the experiment and the CERN accelerator teams for the outstanding performance of the LHC complex.
The ALICE collaboration acknowledges the following funding agencies for their support in building and
running the ALICE detector:
State Committee of Science,  World Federation of Scientists (WFS)
and Swiss Fonds Kidagan, Armenia,
Conselho Nacional de Desenvolvimento Cient\'{\i}fico e Tecnol\'{o}gico (CNPq), Financiadora de Estudos e Projetos (FINEP),
Funda\c{c}\~{a}o de Amparo \`{a} Pesquisa do Estado de S\~{a}o Paulo (FAPESP);
National Natural Science Foundation of China (NSFC), the Chinese Ministry of Education (CMOE)
and the Ministry of Science and Technology of China (MSTC);
Ministry of Education and Youth of the Czech Republic;
Danish Natural Science Research Council, the Carlsberg Foundation and the Danish National Research Foundation;
The European Research Council under the European Community's Seventh Framework Programme;
Helsinki Institute of Physics and the Academy of Finland;
French CNRS-IN2P3, the `Region Pays de Loire', `Region Alsace', `Region Auvergne' and CEA, France;
German BMBF and the Helmholtz Association;
General Secretariat for Research and Technology, Ministry of
Development, Greece;
Hungarian OTKA and National Office for Research and Technology (NKTH);
Department of Atomic Energy and Department of Science and Technology of the Government of India;
Istituto Nazionale di Fisica Nucleare (INFN) and Centro Fermi -
Museo Storico della Fisica e Centro Studi e Ricerche "Enrico
Fermi", Italy;
MEXT Grant-in-Aid for Specially Promoted Research, Ja\-pan;
Joint Institute for Nuclear Research, Dubna;
National Research Foundation of Korea (NRF);
CONACYT, DGAPA, M\'{e}xico, ALFA-EC and the EPLANET Program
(European Particle Physics Latin American Network)
Stichting voor Fundamenteel Onderzoek der Materie (FOM) and the Nederlandse Organisatie voor Wetenschappelijk Onderzoek (NWO), Netherlands;
Research Council of Norway (NFR);
Polish Ministry of Science and Higher Education;
National Authority for Scientific Research - NASR (Autoritatea Na\c{t}ional\u{a} pentru Cercetare \c{S}tiin\c{t}ific\u{a} - ANCS);
Ministry of Education and Science of Russian Federation, Russian
Academy of Sciences, Russian Federal Agency of Atomic Energy,
Russian Federal Agency for Science and Innovations and The Russian
Foundation for Basic Research;
Ministry of Education of Slovakia;
Department of Science and Technology, South Africa;
CIEMAT, EELA, Ministerio de Econom\'{i}a y Competitividad (MINECO) of Spain, Xunta de Galicia (Conseller\'{\i}a de Educaci\'{o}n),
CEA\-DEN, Cubaenerg\'{\i}a, Cuba, and IAEA (International Atomic Energy Agency);
Swedish Research Council (VR) and Knut $\&$ Alice Wallenberg
Foundation (KAW);
Ukraine Ministry of Education and Science;
United Kingdom Science and Technology Facilities Council (STFC);
The United States Department of Energy, the United States National
Science Foundation, the State of Texas, and the State of Ohio.
\end{acknowledgement}

\newpage
\appendix
\section{The ALICE Collaboration}

\bigskip 

B.~Abelev$^{\rm 69}$, 
J.~Adam$^{\rm 37}$, 
D.~Adamov\'{a}$^{\rm 77}$, 
M.M.~Aggarwal$^{\rm 81}$, 
M.~Agnello$^{\rm 105}$$^{\rm ,88}$, 
A.~Agostinelli$^{\rm 26}$, 
N.~Agrawal$^{\rm 44}$, 
Z.~Ahammed$^{\rm 124}$, 
N.~Ahmad$^{\rm 18}$, 
I.~Ahmed$^{\rm 15}$
% ...

\bigskip

\bigskip 

\textbf{\Large Affiliation Notes}

\bigskip 

% ...
$^{\rm VIII}$ Also at: University of Kansas, Lawrence, KS, United States 

\bigskip

\bigskip 

\textbf{\Large Collaboration Institutes}

\bigskip 

% ...
$^{15}$ COMSATS Institute of Information Technology (CIIT), Islamabad, Pakistan\\
% ...
$^{18}$ Department of Physics, Aligarh Muslim University, Aligarh, India\\
% ...
$^{26}$ Dipartimento di Fisica e Astronomia dell'Universit\`{a} and Sezione INFN, Bologna, Italy\\
% ...
$^{37}$ Faculty of Nuclear Sciences and Physical Engineering, Czech Technical University in Prague, Prague, Czech Republic\\
% ...
$^{44}$ Indian Institute of Technology Bombay (IIT), Mumbai, India\\
% ...
$^{69}$ Lawrence Livermore National Laboratory, Livermore, CA, United States\\
% ...
$^{77}$ Nuclear Physics Institute, Academy of Sciences of the Czech Republic, \v{R}e\v{z} u Prahy, Czech Republic\\
% ...
$^{81}$ Physics Department, Panjab University, Chandigarh, India\\
% ...
$^{88}$ Politecnico di Torino, Turin, Italy\\
% ...
$^{105}$ Sezione INFN, Turin, Italy\\
% ...
$^{124}$ Variable Energy Cyclotron Centre, Kolkata, India\\
% ...

\bigskip 

\end{document}